\documentclass[11pt]{article}
\usepackage{a4}
\usepackage[pdftex]{graphicx}
\usepackage[latin1]{inputenc}
\usepackage{ifthen}
\begin{document}

\newcommand{\scale}{0.45}

\newboolean{physik}  
\setboolean{physik}{true}

\ifthenelse{\boolean{physik}}
{
\newcommand{\spin}{spin}
\newcommand{\spinEND}{spin}
\newcommand{\spins}{spins}
\newcommand{\spinss}{spins}
\newcommand{\spinsn}{spins}
\newcommand{\energie}{energy}
\newcommand{\energies}{energies}
\newcommand{\energieForts}{energy}
\newcommand{\gut }{$-1$ }
\newcommand{\schlecht }{$+1$ }
\newcommand{\negativ}{negative}
\newcommand{\positiv}{positive}
\newcommand{\ferromagnetische}{ferromagnetic}
\newcommand{\antiferromagnetische}{antiferromagnetic}
\newcommand{\an}{an}
}
{
\newcommand{\spin}{guest}
\newcommand{\spinss}{guests}
\newcommand{\spinEND}{guest}
\newcommand{\spins}{guests}
\newcommand{\spinsn}{guests}
\newcommand{\energie}{satisfaction}
\newcommand{\energies}{satisfactions}
\newcommand{\energieForts}{satisfaction}
\newcommand{\gut }{$+1$ }
\newcommand{\schlecht }{$-1$ }
\newcommand{\negativ}{positive}
\newcommand{\positiv}{negative}
\newcommand{\ferromagnetische}{friendly}
\newcommand{\antiferromagnetische}{hostile} 
\newcommand{\an}{a}
}

\ifthenelse{\boolean{physik}}
{
\title{Spin glasses: the game}

}
{
\title{My Party !}
}
\author{Alexander K. Hartmann}
\date{}
\maketitle

\ifthenelse{\boolean{physik}}
{
\begin{abstract}
This document presents the rules of a tactical two-player boardgame which
is inspired by spin glasses. The aim is, while placing bonds and spins,
to achieve a majority of the spins facing the chosen direction of each
player. The game has been already successfully used in university teaching
but should be accessible to players from age 10 and up. (Note that
there is a non-physics version of the rules available at 
{\tt www.compphys.uni-oldenburg.de} 
which formulates the game in terms of a comptetion between party hosts.)
Material is included such that a cheap version of the board game
can be made using a color printer, scissors and glue. 
\end{abstract}
}
{}

\begin{center}
\includegraphics[width=0.8\textwidth]{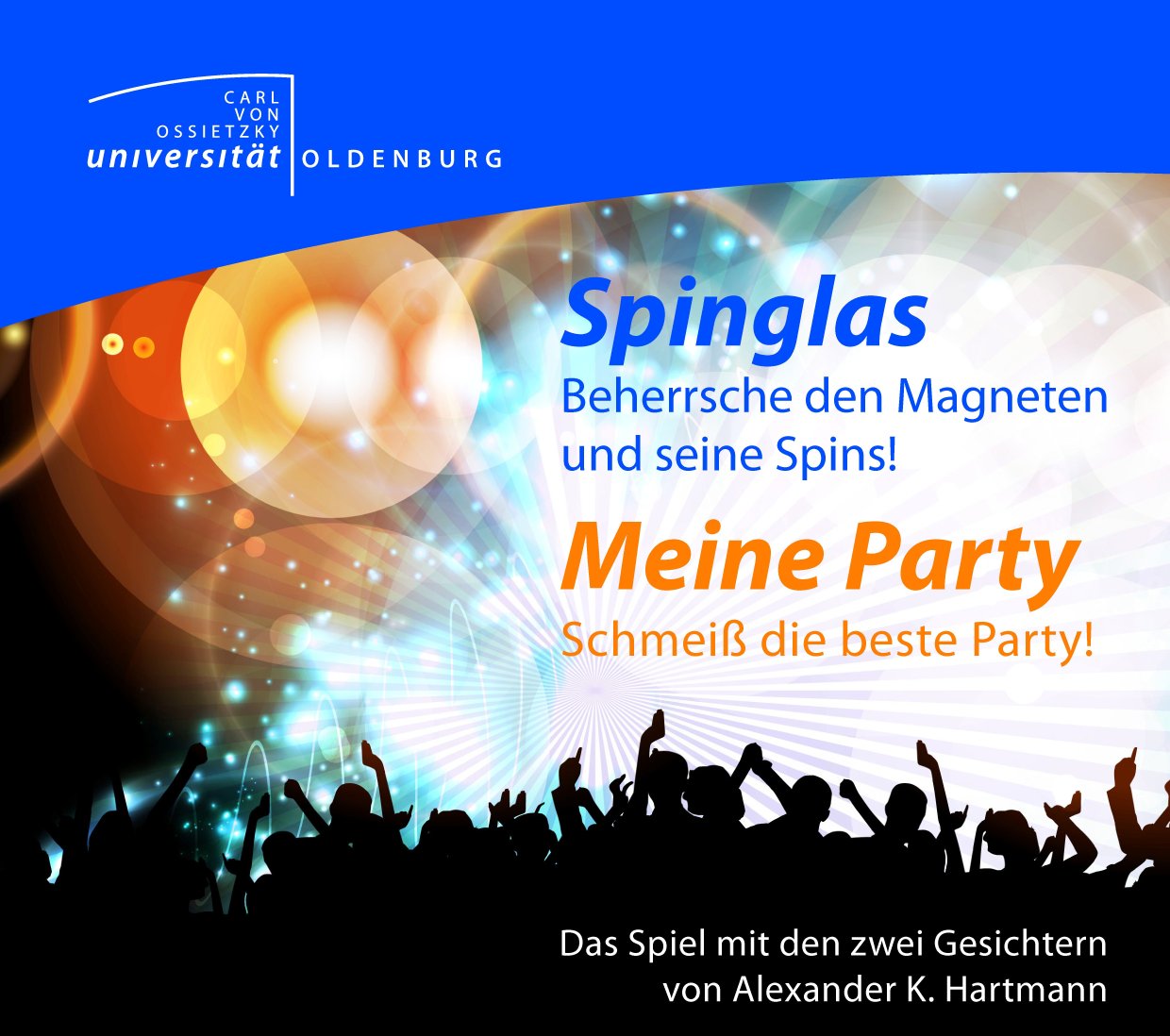}
\end{center}

\ifthenelse{\boolean{physik}}
{
\section{Physical Background}

Spin glasses are disordered magnetic alloys, for example iron-gold,
which exhibit close to absolute zero temperature unusual properties.
These materials, e.g., can remember their ``magnetic history'' they
experienced at higher temperature, altough they seem not to be measurable
at lower temperatures. The most important ingredient of spin glasses
is that they exhibit \emph{ferromagnetic} as well as 
\emph{antiferrmagnetic} interactions. A ferromagnetic interaction
connects spins such that equal orientations of the spins exhibit
lower energies. On contrary, for antiferromagnetic interactions,
opposite orientations of the spins are energetically favorable.
Note that for all physical systems the basic principle holds
that at low temperatures they converge to states of low energy.

Although thousands of scientists have worked on spin glasses during the
past few decades and published their results in about 10000 scientific
papers, still many fundamental properties of spin glasses are poorly 
understood.

\begin{center}
\begin{minipage}[t]{0.9\textwidth}
{\bf Play the spin-glass game and learn to understand the principles of
spin glasses, such that one day you can contribute to uncover
the last secrets of spin glasses!}
\end{minipage}
\end{center}

}
{
\section{Background story}
You are running a pub. To make your pub more fashionable, you are
throwing the ultimate party. Clearly, you want that it will be
a big success by attracting many people to the party. Unfortunately,
your college from the opposite side of the street had the same idea
and is inviting to another party on the same day. You have to
outperform him, more people should come to your party!
For this purpose you should take advantage of the fact that some
potential guests like each other while some others do not. Thus, if you
compose your group of guests in a suitable way, such that their
relations are very friendly, your party will be a big success.
Within this game you can even influence the relations among the
guests, whether they like each other or whether they hate each other.
Hence, it is up to you to achieve a triumphe over your neighbor.

\begin{center}
\begin{minipage}[t]{0.9\textwidth}
{\bf Throw THE party of the year!}
\end{minipage}
\end{center}

}

\section{Game Material}

The game contains the following items:

\begin{itemize}
\item These intructions
\item Two different boards (which can be combined to make one big board)

\item Two types of \emph{pieces}:

\begin{itemize}
\item 
\begin{minipage}[t]{0.5\textwidth}
40 {\em \spins}{} (two-coloured wooded discs with one white and one
black side)
\end{minipage}
\hspace*{5mm}
\begin{minipage}[t]{0.15\textwidth}
\vspace*{-3mm}

\includegraphics[scale=\scale]{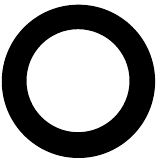}
\includegraphics[scale=\scale]{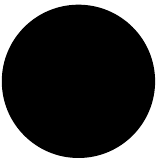}
\end{minipage}

\item 
\begin{minipage}[t]{0.5\textwidth}
60 {\em bonds} (coloured wooden sticks) for
\emph{\ferromagnetische} (blue, 40 pieces) and \emph{\antiferromagnetische} 
(red, 20 pieces) interactions
\end{minipage}
\hspace*{4mm}
\begin{minipage}[t]{0.15\textwidth}
\vspace*{-1mm}

\includegraphics[scale=\scale]{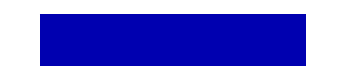}

\includegraphics[scale=\scale]{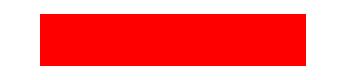}
\end{minipage}
\end{itemize}

\item one little sack 
\item two sets of each 6 {\em action cards} 
\item short instruction (in German) consisting of 4 cards
\item eight additional cards with further information.
\end{itemize}

\section{Aim of the game}

The game is for two players. One player (``player black'') aims
at having as many as possible of the \spins{} being oriented in such a way
that the black side is up. Correspondingly,
the other player (``player white'') tries to maximize the number of
\spins{} with an orientation such that the white side is up.

\section{Game preparation}

One of the boards is selected. The \spins{} are placed next to
the board, they form the \emph{pool}. 
The bonds are put into the sack. Each player receives a
complete set of 6 action cards, selects secretely 3 of them,
and places them in front of him/her upside down, i.e., with the information
side looking down such that it is not visible. 
The three other cards, respectively, are put concealed into the box.

\section{The board}

\begin{minipage}[t]{0.55\textwidth}
Both boards contain \emph{sites} (circles) where during the games the
\spins{} are placed. The sites are connected by \emph{links} where
during the game the \emph{bonds} are placed. The number of adjacent
links per site is varying on the boards.

At the boundary of the boards there are some ``half'' links, which are
only used if both boards are joined to form a large board.

\end{minipage}
\hfill
\begin{minipage}[t]{0.35\textwidth}
\vspace*{-3mm}

\includegraphics[scale=\scale]{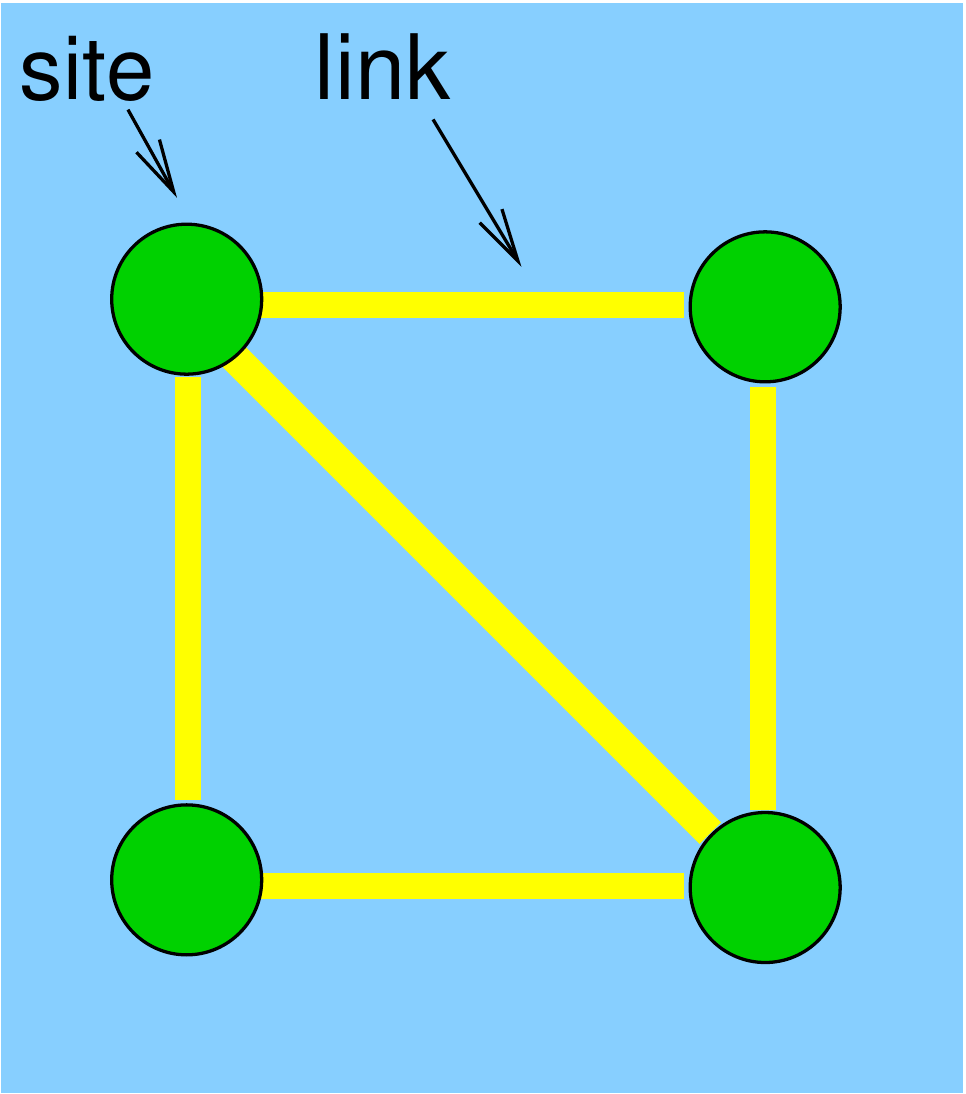}
\end{minipage}

\section{Basic principles/Denominations}

A \spin{} can be placded in two \emph{orientations}, with the color white
up or with the color black up.
\ifthenelse{\boolean{physik}}
{}{In this way it is indicated which party the guest wants to attend.}

\begin{minipage}[t]{0.6\textwidth}
Two \spins{} which are connected by a \ferromagnetische{} bond
prefer to exhibit the same orientation (white/white or black/black).
In the same way two \spins{} joined by \an{} \antiferromagnetische{}
bond prefer to take different orientations. In these two cases, one says
the bond is \emph{satisfied}. Then \an{} \emph{\energie{}} of \gut is
assigned to the bond.

\end{minipage}
\hfill
\begin{minipage}[t]{0.3\textwidth}
\vspace*{-0mm}

\includegraphics[scale=\scale]{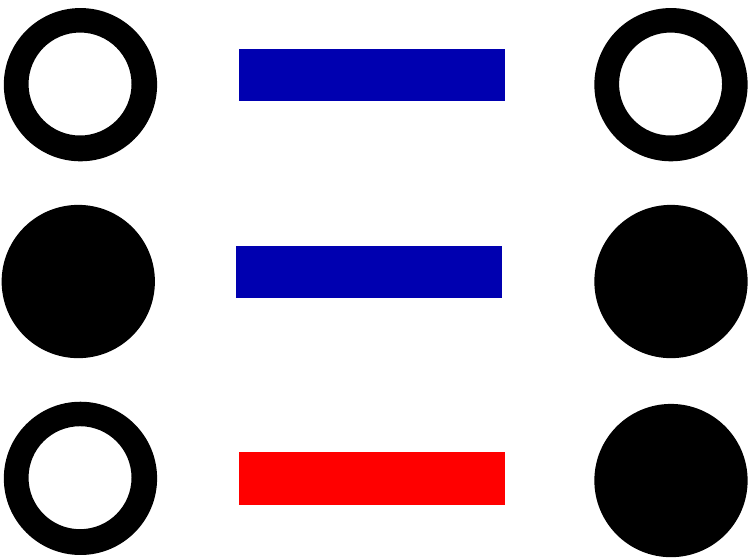}
\end{minipage}

\vspace*{3mm}

\begin{minipage}[t]{0.6\textwidth}
In the opposite case a bond is called \emph{unsatisfied}, i.e.,
if a \ferromagnetische{} bond connects two \spins{} of different
orientations or if \an{} \antiferromagnetische{} bond connects
two \spins{} of equal orientation. To this bond \an{} \energie{}
of \schlecht is assigned.

\end{minipage}
\hfill
\begin{minipage}[t]{0.3\textwidth}
\vspace*{-3mm}

\includegraphics[scale=\scale]{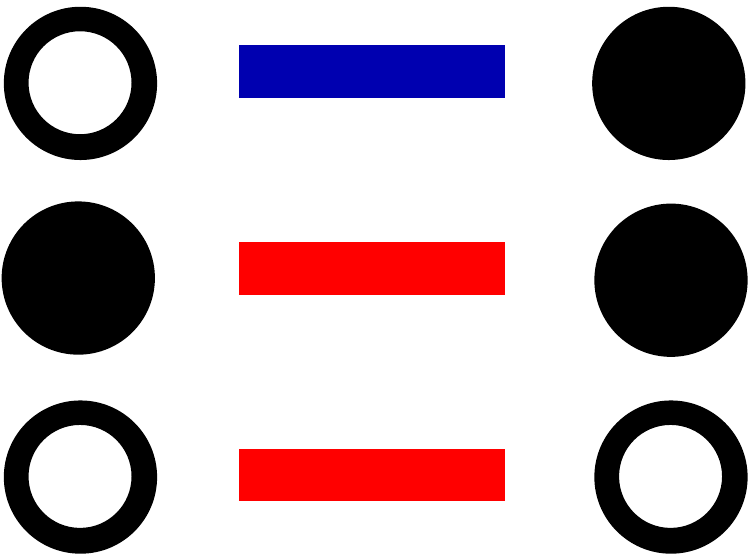}
\end{minipage}

\vspace*{3mm}

Remark: There are situations where it is impossible to satisfy
all bonds, e.g., if three \spins{} interact mutually (forming a triangle)
via \antiferromagnetische{} bonds.

\vspace*{1mm}

Bonds which are not adjacent to \emph{two} \spins{} are \emph{neutral}
(\energie 0).

\vspace*{1mm}

The most important quantity for each \spin{} is the total \energie{}
of all bonds adjacent to the \spin, i.e., the sum of the \energie{}
values. Links where (yet) no bonds are placed and also neutral bonds
(where at least at one side no \spin{} is placed) are not taken
into account when calculating the total \energie{}.

A \spin{} is called \emph{stable}, if for its adjacent bonds there 
are \emph{more} satisfied  than unsatisfied  (total \energie{} \negativ).
A \spin{} is called \emph{unstable}, if for its adjacent bonds 
there are \emph{less} satisfied 
than unsatisfied  (total \energie{} \positiv).
A \spin{} is called \emph{free}, if for its adjacent bonds 
the number  of satisfied and
 unsatisfied  is \emph{the same} (total \energie{} zero).
In particular a \spin{} is free, if no bonds are adjacent.

Example: for the \spin{} marked with 'X' we obtain:

\begin{minipage}[t]{0.32\textwidth}
\begin{center}
\includegraphics[scale=\scale]{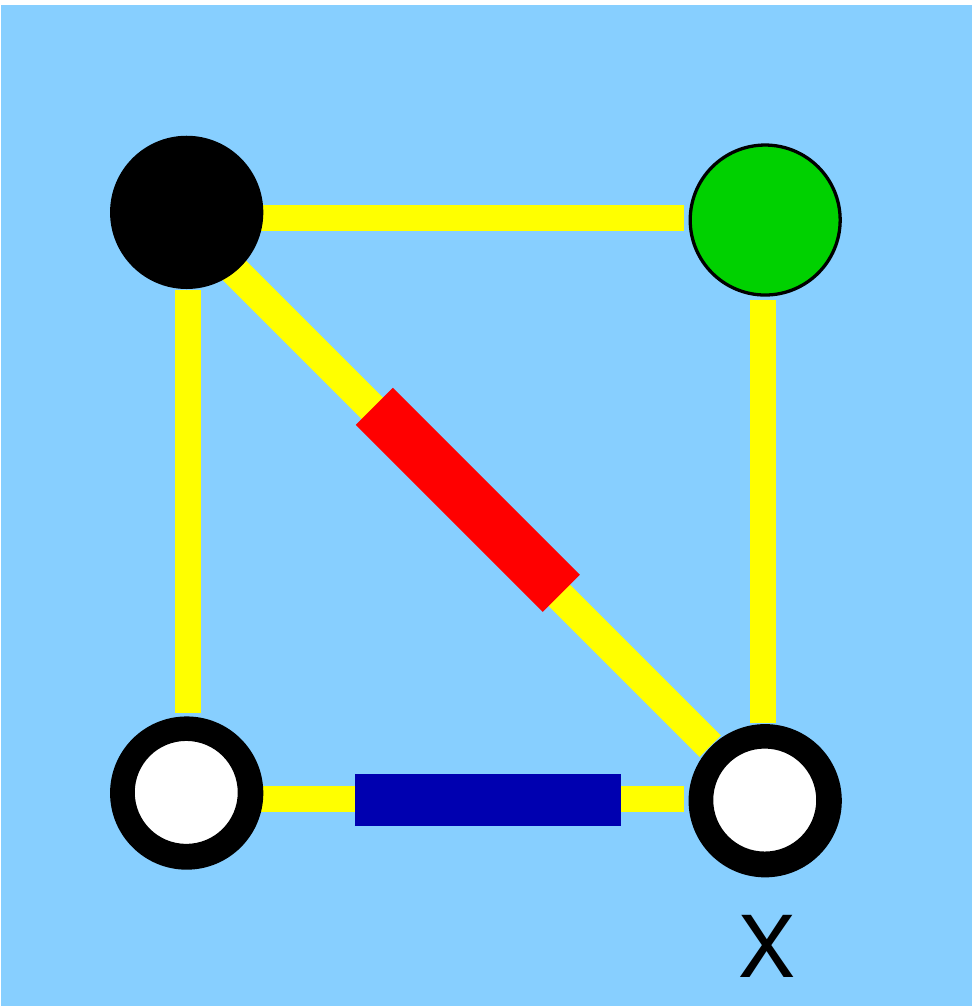}

total \energie{} \ifthenelse{\boolean{physik}}{-2}{2}
\end{center}
\end{minipage}
\begin{minipage}[t]{0.32\textwidth}
\begin{center}
\includegraphics[scale=\scale]{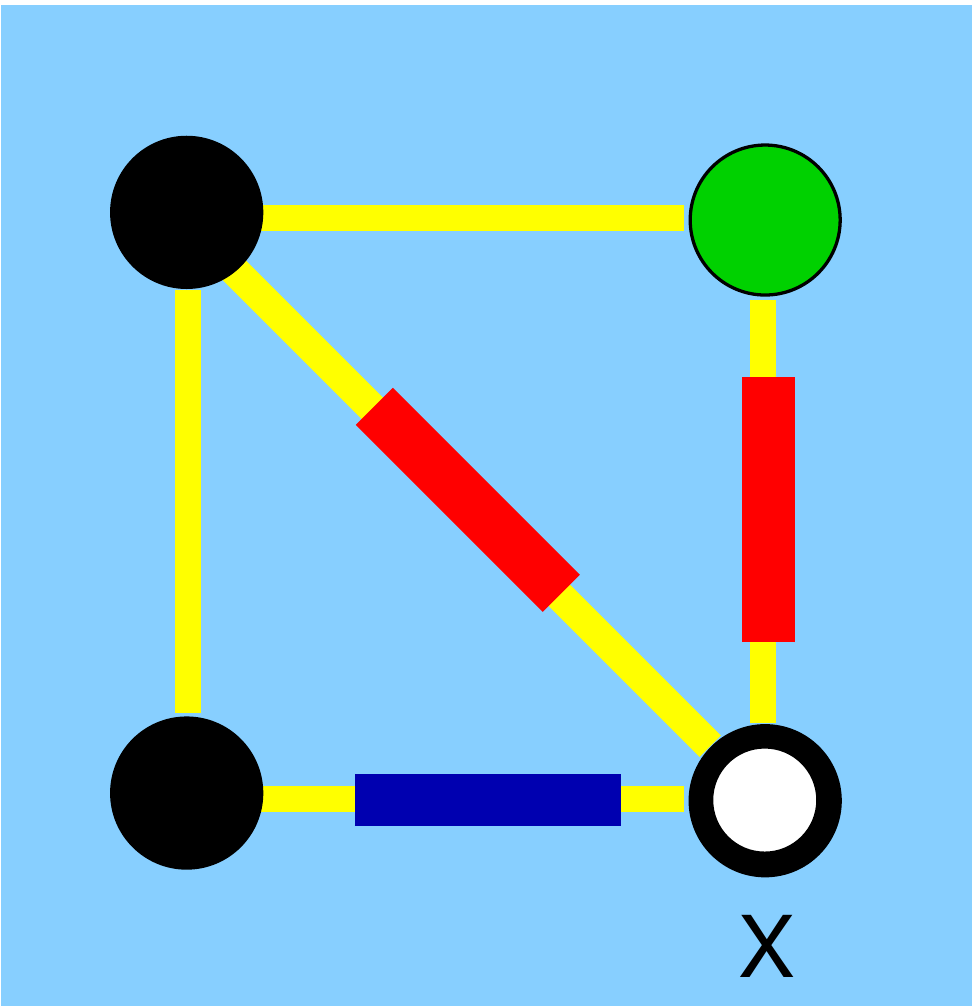}

total \energie{} 0
\end{center}
\end{minipage}
\begin{minipage}[t]{0.32\textwidth}
\begin{center}
\includegraphics[scale=\scale]{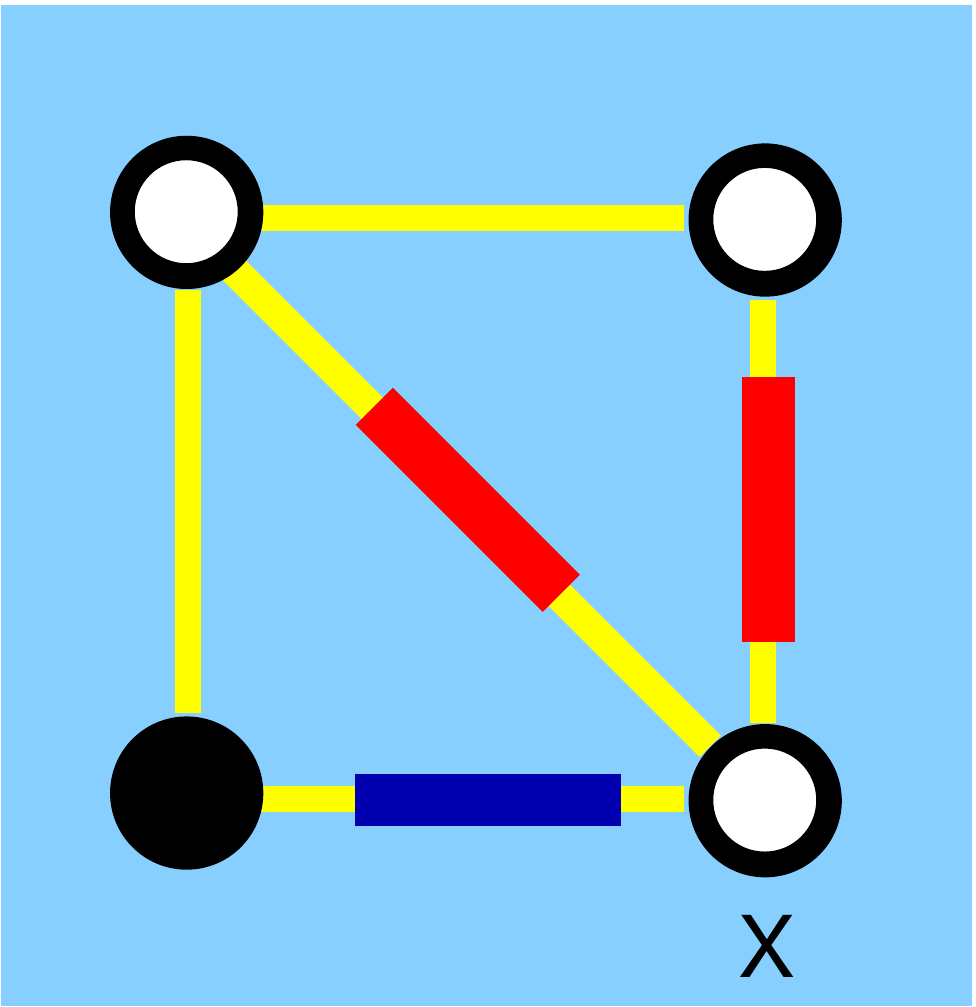}

total \energie{} \ifthenelse{\boolean{physik}}{3}{-3}
\end{center}
\end{minipage}

\section{The game}

The players perform moves alternatingly. The junger one starts
and gets assigned the color white. A move consits of \emph{some}
of the {\bf steps} listed below. Each step can occur once,
several times or zero times within one single move, while
observing the basic rules mentioned here and on the next page:

\begin{itemize}

\item Blindly drawing a bond from the sack.

\item Placing a bond on an unoccupied link, i.e., where so far no
bond was placed, no matter what the
orientation of the adjacent \spins{} are, if there are any.

\item Taking one \spin{} from the pool.

\item Placing a \spin{} on an unoccupied site, i.e., where no
\spin{} was placed so far. The orientation (black/white) \emph{must} 
be chosen such that the total \energie{} of the \spin{} is \emph{NOT
\positiv}, i.e., it may be \negativ{} or zero.

\item Flipping of \spins{} located on the board: a \spin{}
showing black on top before will show white on top afterwards, and
vice versa. It is only allowed to flip \spins{}, where the total
\energie{} (sum of \energies) is \positiv{} or zero, i.e.,
where the \spin{} is unstable or free.

Remark: \spins{} exhibiting a \positiv{} total \energie{} may be flipped
but are not required to be flipped, in contrast to
the strict rule when placing a \spin. This freedom 
holds in particular if first
a \spin{} is placed on a site without adjacent bonds (all adjacent links
are not occupied, leading to total \energie{} zero) and next a bond
is placed adjacent to the \spin. (Also it may just happen and it
is allowed that a player does not spot  a \spin{} which can be flipped, 
or forgets to flip a \spin.)

\end{itemize}

{\bf Basic rules}: For s standard move, \emph{a total} of three pieces
(bonds/\spins) is taken from the sack and/or pool and placed on the board.
This may be three bonds in one move, or one bond and two \spins, etc.
(Exceptions may occur when using action cards, see section
\ref{sec:aktionskarten}.) All pieces taken within a move have to be placed
on the board during the move.

The order of the steps within a move is arbitrary. \emph{Example}:
a player may first draw a bond and then take a \spin{} from the pool.
Next the player places the bond and then the \spin{}, or the other
way round. The player is also allowed to take a third piece first.
During a move one or several \spins{} may be flipped, if allowed,
at any time. Also up to one action card may be used within one move
any time.

A move is {\bf finished} if a player has placed three pieces (\spins/bonds)
and announces that he or she is finished.

{\bf Tactical advice}: Via placing bonds and via flipping \spins{}
it is possible to achieve that adjacent \spins, which were stable
before, become free or even unstable and thus can be flipped.
This is the main mechanism to flip \spins{} from the opponent's
orientation to the own orientation.

With this set of rules, together with the explenations of the
action cards in section \ref{sec:aktionskarten} and with the
rules for the finishing the game  (section \ref{sec:abrechung}), 
the game is completely described. For a better comprehendsion,
next some examples are given, in particular for illustrating the tactical 
advice.

\section{Example moves}

\begin{minipage}[t]{0.62\textwidth}
Player white starts and draws one bond from the sack and happens
to obtain a \ferromagnetische{} (blue) bond.

The player places the bond on a link. Next, the player takes a \spin{}
from the pool and places it adjacent to the bond. Since the
bond is still neutral, the \spin{} is free and the player can chose the
orientation of the \spin. He/she choses the orientation white.
The player takes another \spin{} and places it on the other end of the bond.
Due to the \ferromagnetische{} bond and due to the white orientation of
the first \spin{}, now only the orientation white is possible. The
situation shown to the right is obtained.
\end{minipage}
\hfill
\begin{minipage}[t]{0.35\textwidth}

\newpage

\includegraphics[scale=\scale]{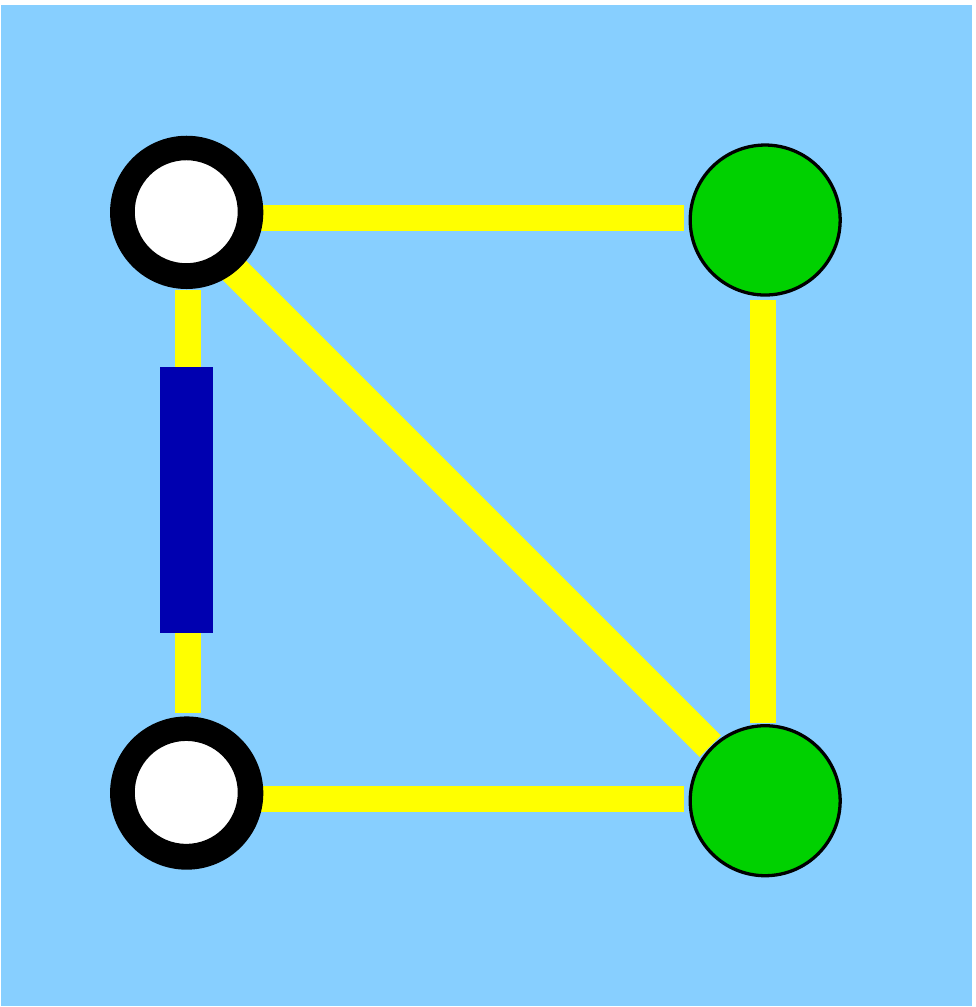}
\end{minipage}

\vspace*{2mm}

\begin{minipage}[t]{0.62\textwidth}
Now it is the turn of player black. He/she draws a bond from the sack
and happens to obtain also a \ferromagnetische{} (blue) bond.
The player decides to draw another bond and obtaines by chance
another \ferromagnetische{} bond. Now the player \emph{first} places
a \spin{} on the site marked by X. He/she is allowed to chose the orientation
black since currently no bonds are adjancent to the site.
\emph{Next}, the player places the two \ferromagnetische{}
bonds between site X and the sites A and B which are already occupied by
\spins. Now the two white \spins{} A and B have become free,
since each of them is adjacent to one satisfied bond (connecting
the \spins{} 
\end{minipage}
\hfill
\begin{minipage}[t]{0.35\textwidth}
\vspace*{-3mm}

\includegraphics[scale=\scale]{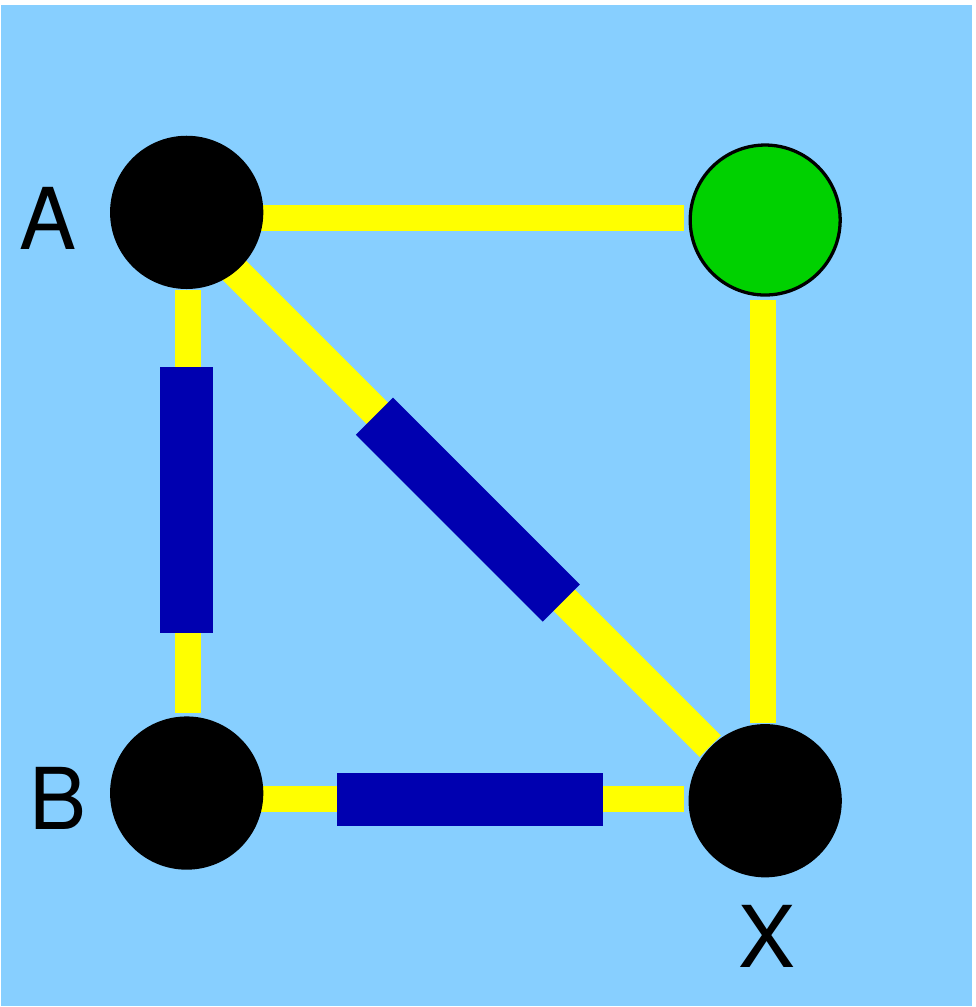}
\end{minipage}

\vspace*{1mm}

A and B) and one unsatisfied bond (adjacent to \spin{} X).
Player black now is allowed to flip any of the two free \spins, e.g., A.
Now \spin{} B has become unstable and can be flipped as well.
This results in the situation shown, where now all \spins{}
are oriented in favor of player black.

Remark: If player black had placed first a bond (or two)
adjacent to the two white \spins{}, he/she would have been
forced to place the \spin{} at site X with orientation white,
because it is not allowed to place a \spin{} with \positiv{} \energie.
Hence, the order of the steps must be chosen carefully.

Note that within the resulting situation the \spins{} are more
stable, since it needs at least two unsatisfied bonds to free and
thus flip a \spin.
As we have seen in the previous example move, pairs (or chains) of \spins{}
can be flipped easily.

\begin{minipage}[t]{0.62\textwidth}
Remark: if player black had obtained not blue but insetad one red
and one blue bond, he/she could have turned the two white \spins{} A
and B as well. In this case the situation shown to the right could
have been obtained. Currently, \spin{} A is stable. Nevertheless,
\spins{} X and B are free since both are adjacent to one satisfied
 and one unsatisfied bond. Thus, player white could in his/her
move flip \spin{} B. Now \spin{} A has been freed, which can be flipped
as well to white. 
Finally, \spin{}
X is free and can be flipped as well. All \spins{} have been flipped
from black to white. Nevertheless, the situation  is still not stable,
player black could
\end{minipage}
\hfill
\begin{minipage}[t]{0.35\textwidth}
\vspace*{-3mm}

\includegraphics[scale=\scale]{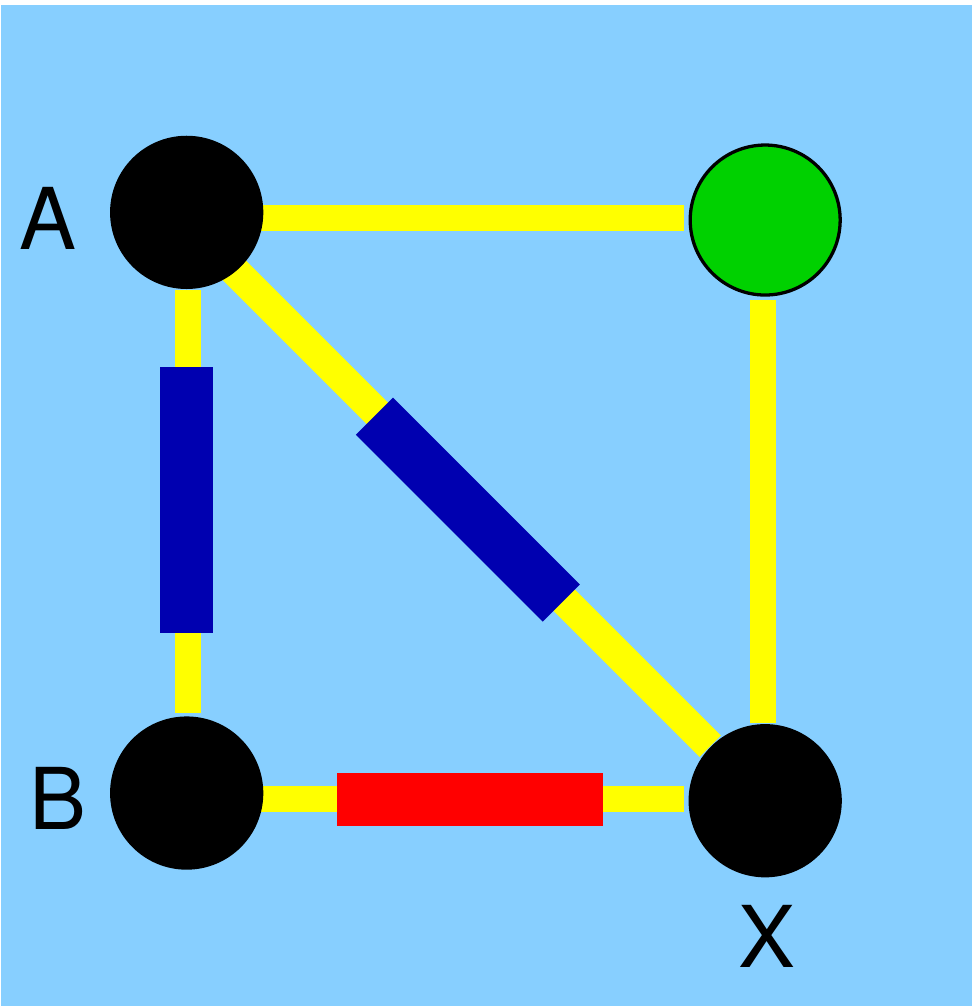}
\end{minipage}

\vspace*{1mm}

 in the same way flip all three \spins{} again.
Therefore, all three \spins{} are directly and indirectly (through
neighboring free \spins) free. Note that such \spins{}
will be ignored in the final evaluation of the game
(see section \ref{sec:abrechung}).

\section{Action cards}
\label{sec:aktionskarten}

At the beginning of the game, each player receives a complete
set of 6 action cards. From this set each player selects three cards,
hidden from the view of the other player. There are the
following action cards which can be used at \emph{any time}
during a move, but \emph{at most one} during one move. After
an action card has been used, it is put aside and cannot be used again
during the same game.

\begin{minipage}[t]{0.2\textwidth}
\vspace*{14mm}

\includegraphics[width=\textwidth]{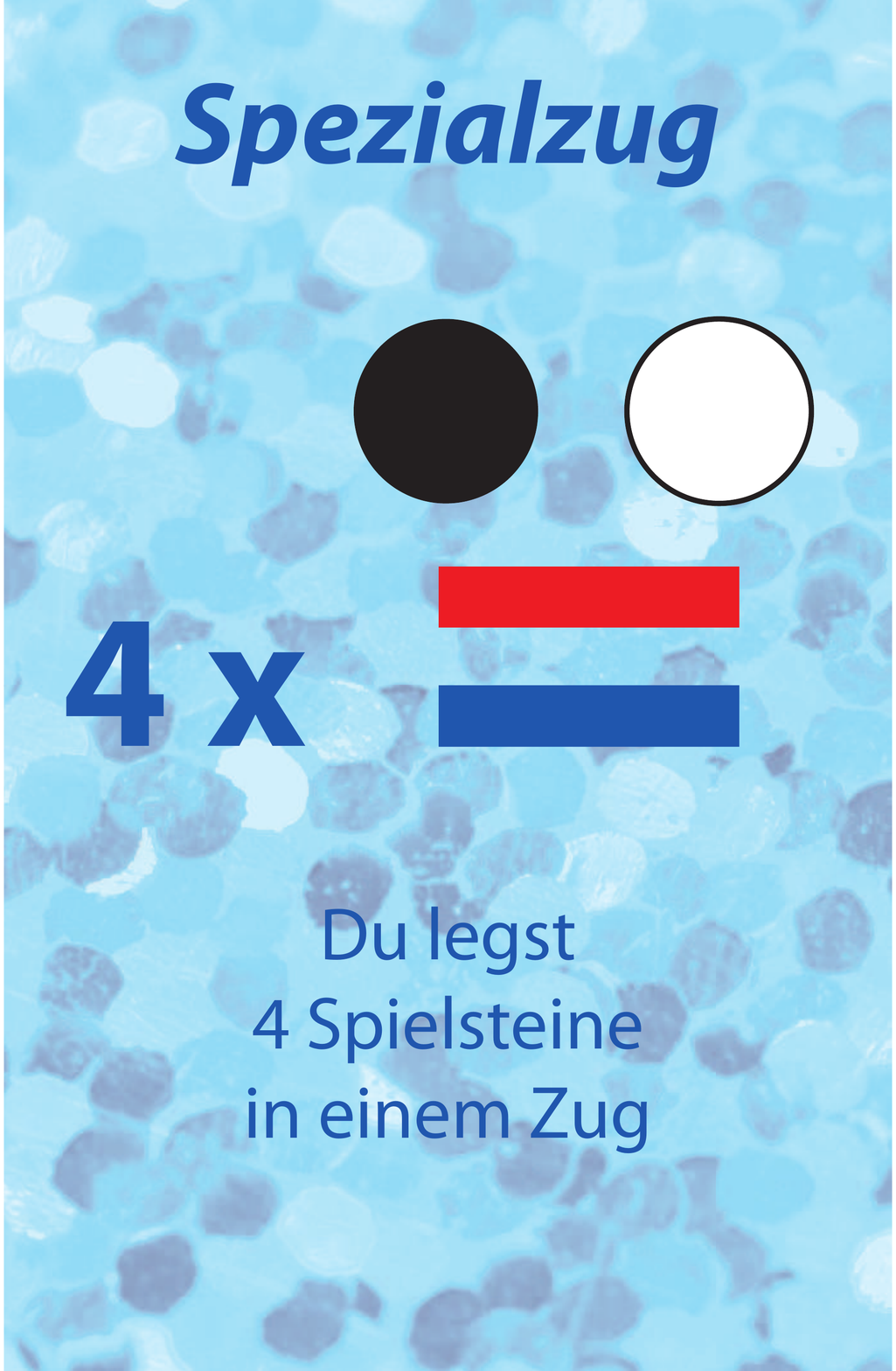}
\end{minipage}
\hspace*{3mm}
\begin{minipage}[t]{0.7\textwidth}
For the ``special move'' card, the player has to take and place 
four pieces instead
of three. Apart from this, the rules for a move do not change.
\end{minipage}

\vspace*{-6mm}

\begin{minipage}[t]{0.2\textwidth}
\vspace*{14mm}

\includegraphics[width=\textwidth]{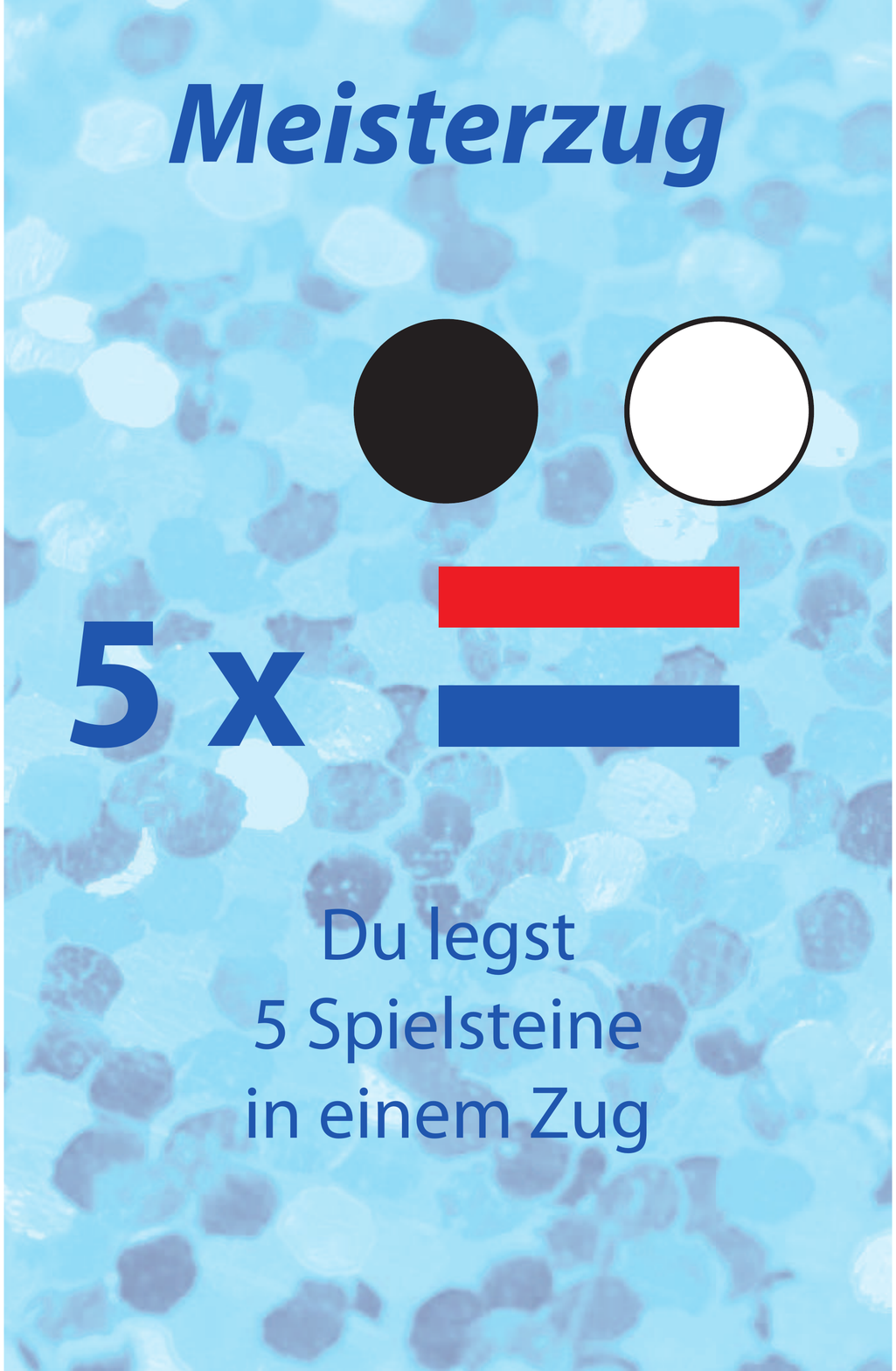}
\end{minipage}
\hspace*{3mm}
\begin{minipage}[t]{0.7\textwidth}
When using the ``master move'' card, the player must take and place
five peices instead of three.
Apart from this, the rules for a move do not change.
\end{minipage}

\vspace*{-6mm}

\begin{minipage}[t]{0.2\textwidth}
\vspace*{14mm}

\includegraphics[width=\textwidth]{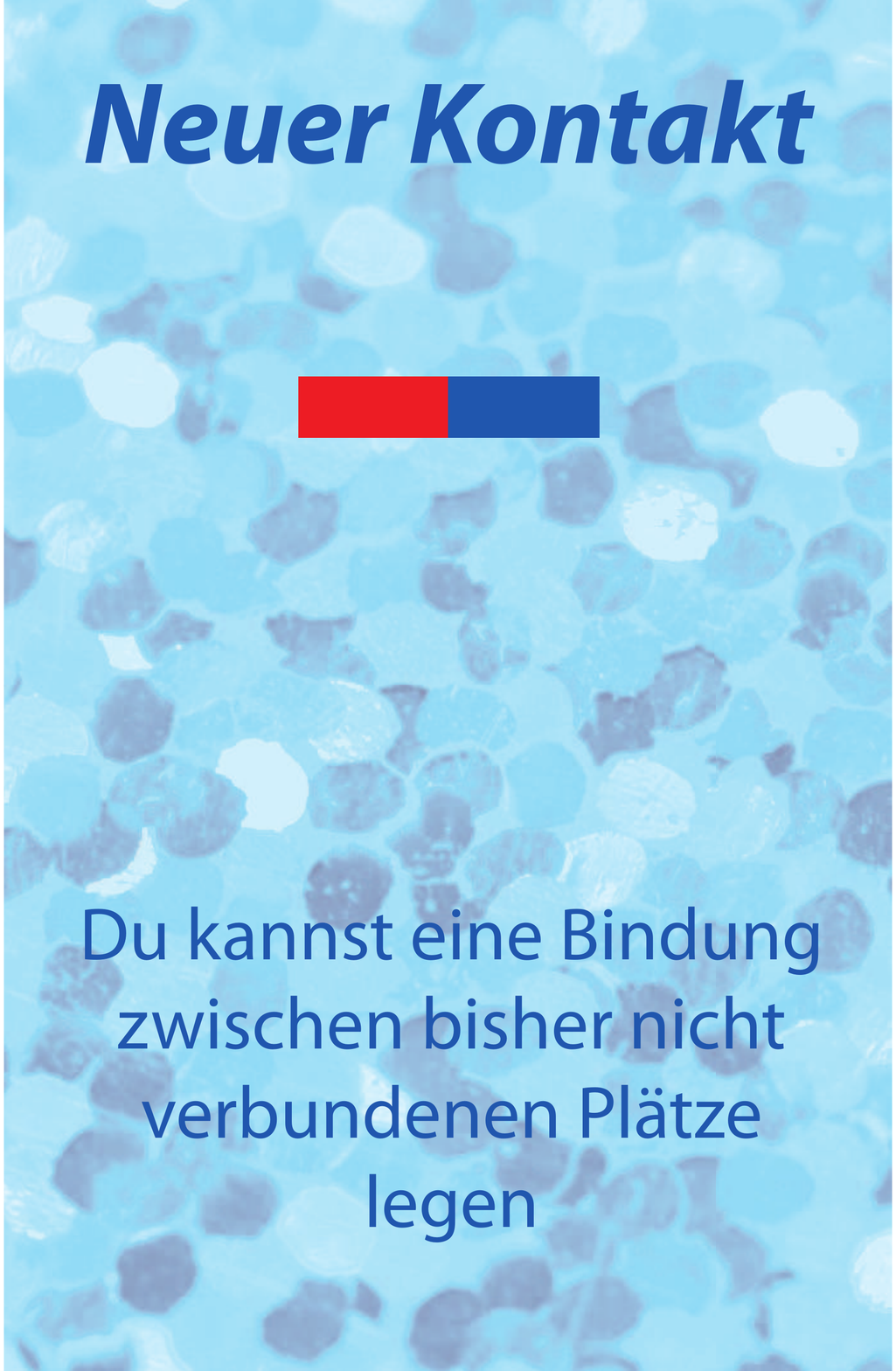}
\end{minipage}
\hspace*{3mm}
\begin{minipage}[t]{0.7\textwidth}
Using the ``new link'' action card,
the player places a bond between two sites which are not
connected by a link. It does not matter whether there are already
\spins{} placed on theses sites. It is like a new link is created
between the two sites, where the bond is placed.
 Requirement: the imaginary straight line between
the two sites must not touch existing links or sites.
Remark: still the total number of pieces placed 
is exactly three wthin the move.
\end{minipage}

\vspace*{-6mm}

\begin{minipage}[t]{0.2\textwidth}
\vspace*{14mm}

\includegraphics[width=\textwidth]{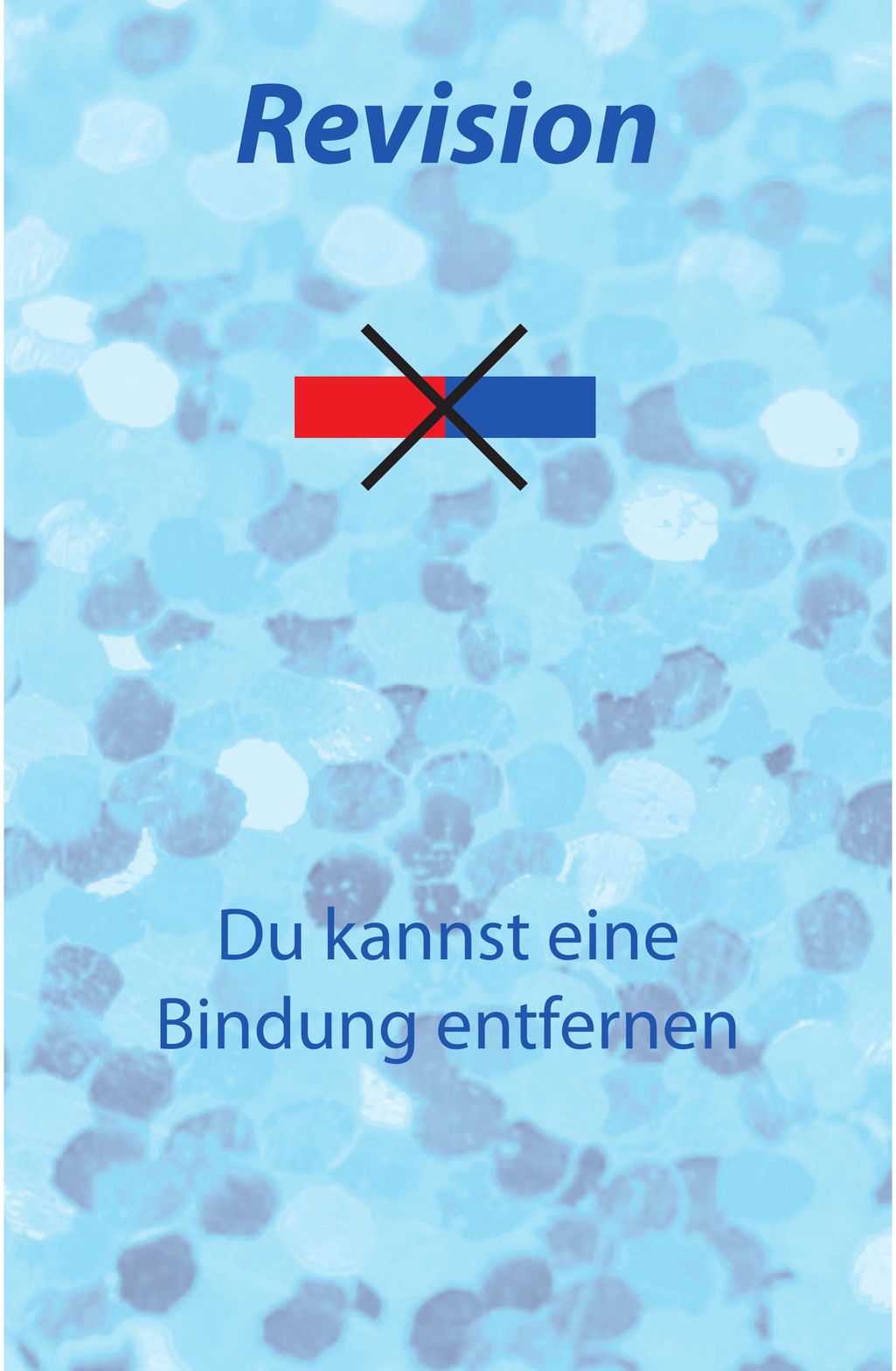}
\end{minipage}
\hspace*{3mm}
\begin{minipage}[t]{0.7\textwidth}
Using the ``revision'' card, 
the player removes an arbitrary bond and puts it back into the sack.\\
Remark 1: if the removed bond was previously placed via
a ``new link'' card, this link is destroyed as well.\\
Remark 2: Via well planned usage of this card, some \spins{} might
become free and thus can be flipped.
\end{minipage}

\newpage

\begin{minipage}[t]{0.2\textwidth}
\vspace*{14mm}

\includegraphics[width=\textwidth]{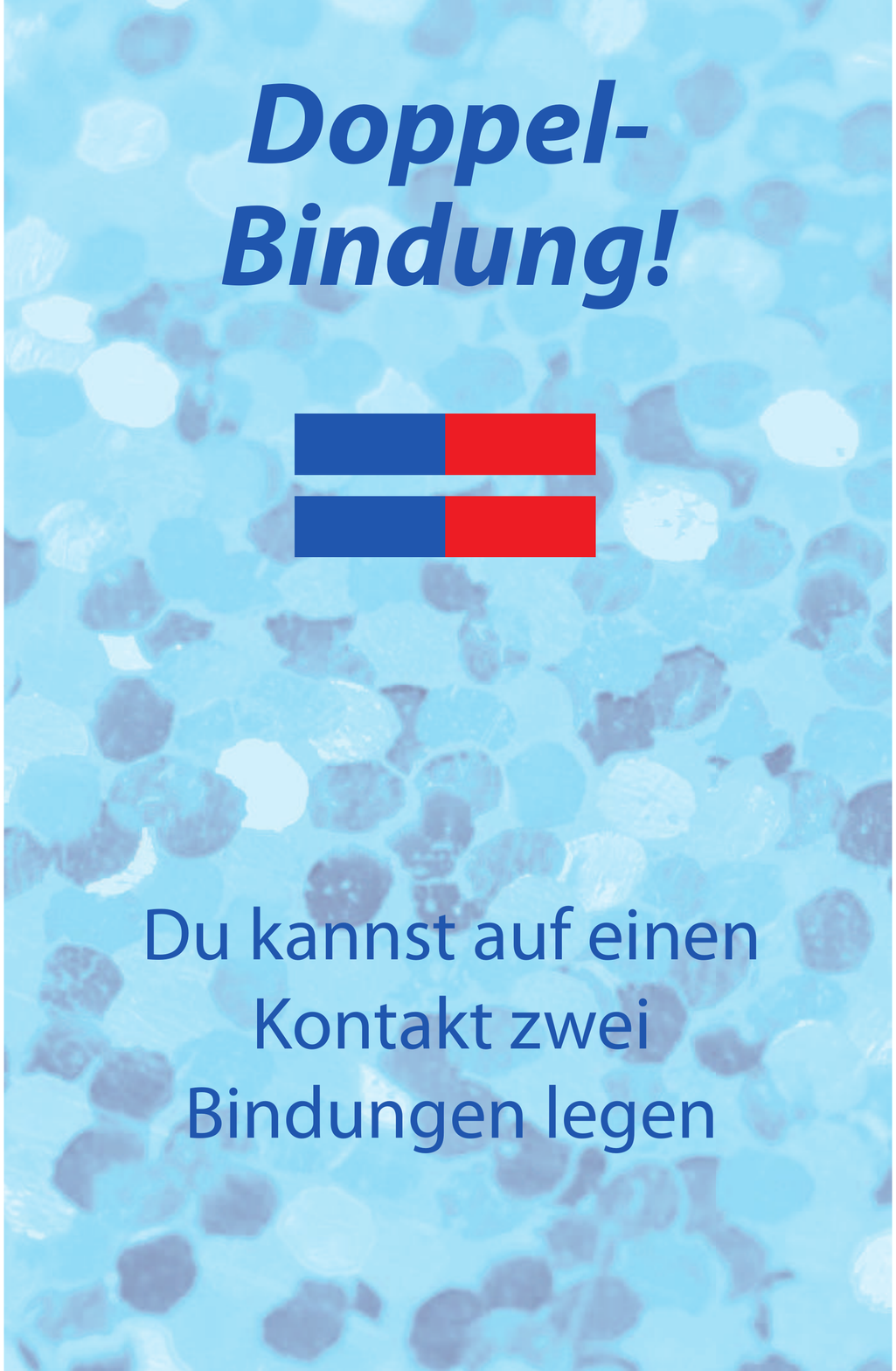}
\end{minipage}
\hspace*{3mm}
\begin{minipage}[t]{0.7\textwidth}
When using the ``double bond'' card, the player puts one of the bonds
played during the move next to an exisiting bond.

Remark: Both bonds contribute to the calculation of the \energie.
Thus, the corresponding link attains twice the normal importance,
if both bonds placed on it are the same. If the two bonds
are different, they cancel each other, i.e., they neutralized
each other.
\end{minipage}

\vspace*{-6mm}

\begin{minipage}[t]{0.2\textwidth}
\vspace*{15mm}

\includegraphics[width=\textwidth]{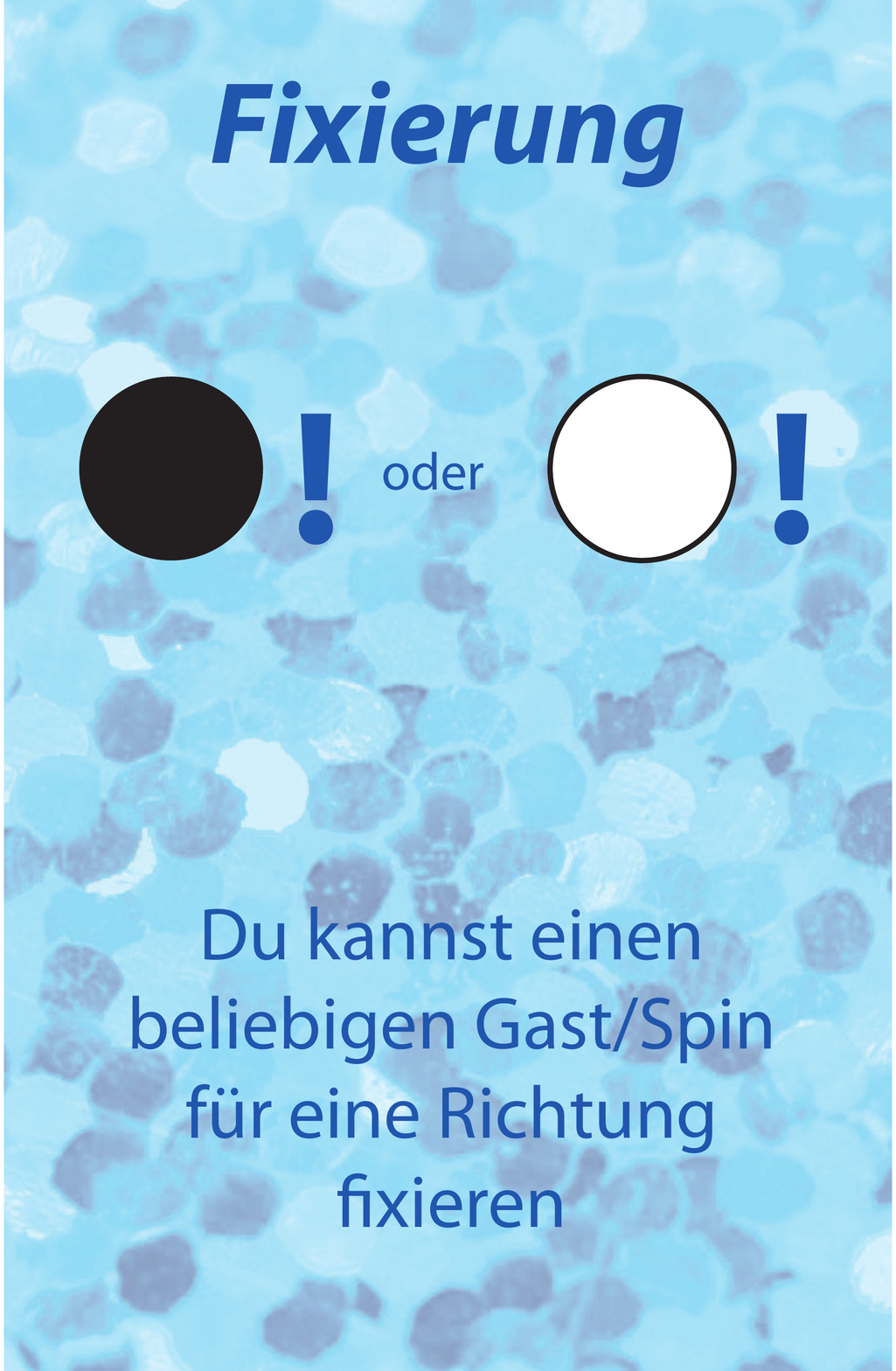}
\end{minipage}
\hspace*{3mm}
\begin{minipage}[t]{0.7\textwidth}
The ``fixing'' card allows a player to orient and fix an arbitrary \spin{} 
on the borad in an arbitrary orientation,
independent from the adjacent bonds and the neighboring \spins.
This \spin{} is marked by putting another \spin{} 
on the top of the \spin. Still, the fixed \spin{}
 is considered and counted as one \spin.
During the remaining game, the fixed \spin{} is not allowed to flip, even not
when using the other ``fixing'' card.

Remark: This is probably the strongest action card. Via
a carfully chosen application, one might create a large cascade
of flipped \spins.
\end{minipage}

\section{Finish}

The game is finished when all sites are occupied by \spins{} and
all links are occupied by bonds. Thus, it might happen that during
the last move only less than three pieces can be placed, if no unoccupied
site or bond is available. In particular, during the last move
a player might not have the choice of the pieces. It might happen,
e.g., that there are only two sites left and no links, thus only
two \spins{} can be placed.

After the final move, no action card can be used, by any player.

\section{Final evaluation}
\label{sec:abrechung}

There might be \spins{} which are free after the game has finsied,
i.e., they may be flipped. In particular, there might be \spins, which
are \emph{indiretly} free, i.e., they may be flipped after a free 
neighboring \spin{} has been fliped. This may lead to cascades of
free \spins.

Remark 1: \spins{} which are fixed are by definition not free.

Remark 2: \spins{} which are adjacent to an odd number of bonds
can never be free.

Remark 3: if during the evaluation unstable \spins{} are detected
(which have been overseen before), any player is allowed to flip them.

Thus, first the total set of directly and indirectly free \spins{}
is identified. They are removed in one strike. This means, one
does not remove a free \spin{} once it is identified, but only after
all free \spins{} have been detected.

After removal of the free \spins, the \spins{} of each orientation
are counted. The player wins who has the majority of \spins{} showing
his/her orientation. If the number of \spins{} showing the
two different orientations is the same, the game ends in a draw.\\[0.5cm]

\section{Ideas for variants}

\begin{itemize} 
\item Both boards may be joined to form a large board. In this case
each player gets assigned all six action cards (they need not
to be kept concealed).

\item The players are not forced to place the pieces immediately. Instead,
they can be (partially) collected such that during a later move more
than three pieces can be placed.

\item One can remove chance completely from the game, if the
bonds are not taken randomly from the sack. Instead, each player
receives the same set of bonds from which he/she may select
some during a move.

\item Pro version: You can divide the game into two phases: First,
only bonds are placed. After all bonds are placed, the \spins{}
are placed using the usual rules. Hence, when putting the bonds
a player has to have already a good plan (which, on the other
hand, may be exploited by the opponent).
\end{itemize}

\vspace*{1cm}

{\Large \bf Enjoy the game! }
\hspace*{3cm}\\[0.5cm]

\hspace*{0.5cm} Alexander K. Hartmann

\vfill
{\small The author thanks for various support: all
test players, the BIS publisher University of Oldenburg
and in particular Hans-Joachim W\"atjen and Hille Schulte, 
the University society
of Oldenburg, 
Lothar Witt, Prof.\ Dr.\ Martin Holthaus, Prof.\ Dr.\ Michael Komorek,
Prof.\ Dr.\ Christoph Linau, Prof.\ Dr.\ Jürgen Parisi, Prof.\ Dr.\
Joachim Peinke, and Prof.\ Dr.\ Björn Poppe.}

\newpage

{\Large \bf Appendix}\\

For trying the game, you might want to print it and cut the pieces out,
see the following pages.
Note that you can buy the professionally produced 
game for 14.50 Euros  (which is slightly below the production
costs due to support from the University!) 
plus pp at the University of Oldenburg 
via emailing to \verb!bisverlag@uni-oldenburg.de!.

\begin{center}
\includegraphics[width=0.99\textwidth]{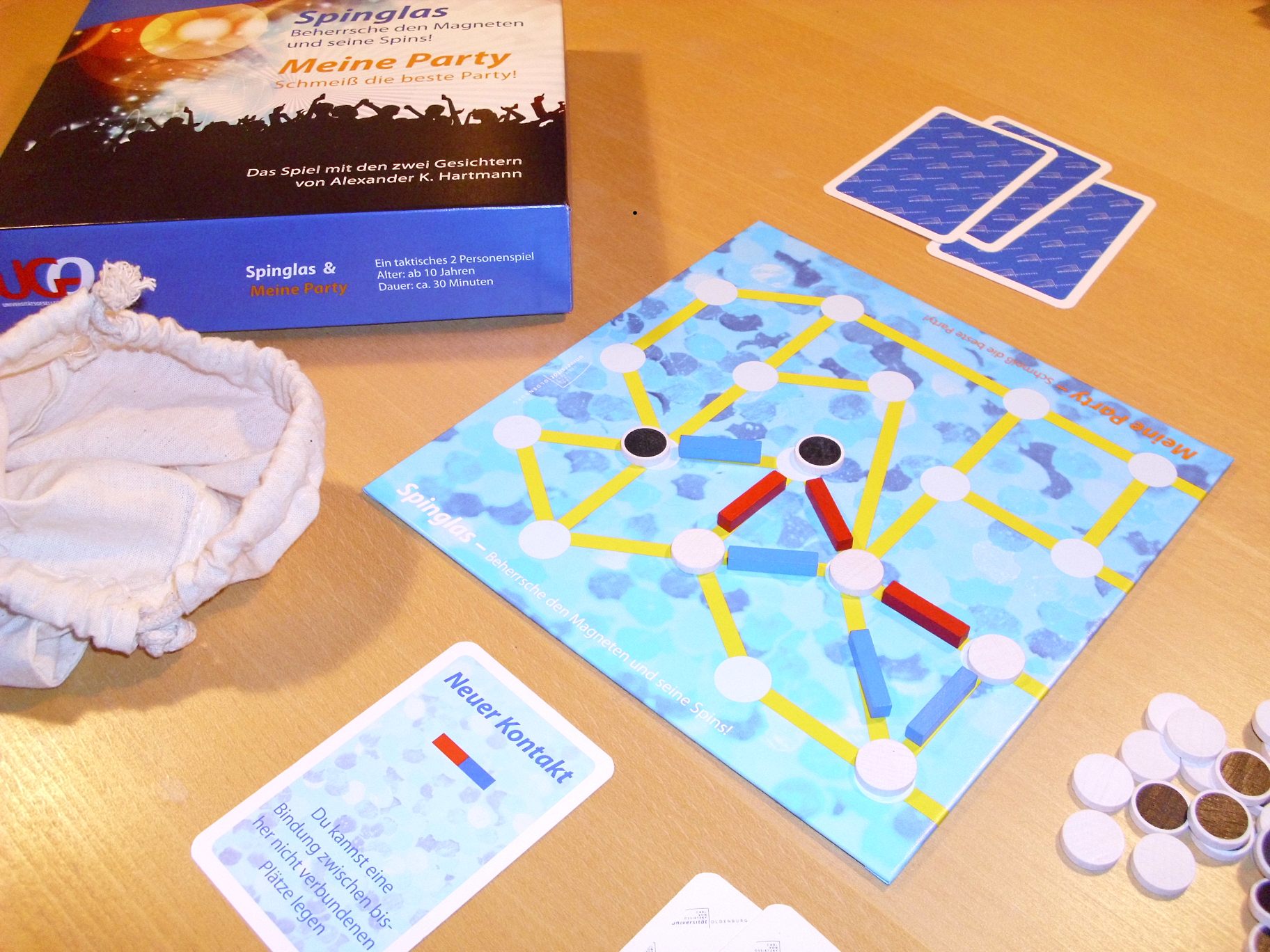}
\end{center}

\newpage

We suggest to print the \spins{} below and glue them on coins,
one side black, one side white.\\

\includegraphics[width=15mm]{spin_weiss}
\includegraphics[width=15mm]{spin_weiss}
\includegraphics[width=15mm]{spin_weiss}
\includegraphics[width=15mm]{spin_weiss}
\includegraphics[width=15mm]{spin_weiss}
\includegraphics[width=15mm]{spin_weiss}
\includegraphics[width=15mm]{spin_weiss}
\includegraphics[width=15mm]{spin_weiss}
\includegraphics[width=15mm]{spin_weiss}
\includegraphics[width=15mm]{spin_weiss}

\includegraphics[width=15mm]{spin_weiss}
\includegraphics[width=15mm]{spin_weiss}
\includegraphics[width=15mm]{spin_weiss}
\includegraphics[width=15mm]{spin_weiss}
\includegraphics[width=15mm]{spin_weiss}
\includegraphics[width=15mm]{spin_weiss}
\includegraphics[width=15mm]{spin_weiss}
\includegraphics[width=15mm]{spin_weiss}
\includegraphics[width=15mm]{spin_weiss}
\includegraphics[width=15mm]{spin_weiss}

\includegraphics[width=15mm]{spin_weiss}
\includegraphics[width=15mm]{spin_weiss}
\includegraphics[width=15mm]{spin_weiss}
\includegraphics[width=15mm]{spin_weiss}
\includegraphics[width=15mm]{spin_weiss}
\includegraphics[width=15mm]{spin_weiss}
\includegraphics[width=15mm]{spin_weiss}
\includegraphics[width=15mm]{spin_weiss}
\includegraphics[width=15mm]{spin_weiss}
\includegraphics[width=15mm]{spin_weiss}

\includegraphics[width=15mm]{spin_weiss}
\includegraphics[width=15mm]{spin_weiss}
\includegraphics[width=15mm]{spin_weiss}
\includegraphics[width=15mm]{spin_weiss}
\includegraphics[width=15mm]{spin_weiss}
\includegraphics[width=15mm]{spin_weiss}
\includegraphics[width=15mm]{spin_weiss}
\includegraphics[width=15mm]{spin_weiss}
\includegraphics[width=15mm]{spin_weiss}
\includegraphics[width=15mm]{spin_weiss}

\includegraphics[width=15mm]{spin_schwarz}
\includegraphics[width=15mm]{spin_schwarz}
\includegraphics[width=15mm]{spin_schwarz}
\includegraphics[width=15mm]{spin_schwarz}
\includegraphics[width=15mm]{spin_schwarz}
\includegraphics[width=15mm]{spin_schwarz}
\includegraphics[width=15mm]{spin_schwarz}
\includegraphics[width=15mm]{spin_schwarz}
\includegraphics[width=15mm]{spin_schwarz}
\includegraphics[width=15mm]{spin_schwarz}

\includegraphics[width=15mm]{spin_schwarz}
\includegraphics[width=15mm]{spin_schwarz}
\includegraphics[width=15mm]{spin_schwarz}
\includegraphics[width=15mm]{spin_schwarz}
\includegraphics[width=15mm]{spin_schwarz}
\includegraphics[width=15mm]{spin_schwarz}
\includegraphics[width=15mm]{spin_schwarz}
\includegraphics[width=15mm]{spin_schwarz}
\includegraphics[width=15mm]{spin_schwarz}
\includegraphics[width=15mm]{spin_schwarz}

\includegraphics[width=15mm]{spin_schwarz}
\includegraphics[width=15mm]{spin_schwarz}
\includegraphics[width=15mm]{spin_schwarz}
\includegraphics[width=15mm]{spin_schwarz}
\includegraphics[width=15mm]{spin_schwarz}
\includegraphics[width=15mm]{spin_schwarz}
\includegraphics[width=15mm]{spin_schwarz}
\includegraphics[width=15mm]{spin_schwarz}
\includegraphics[width=15mm]{spin_schwarz}
\includegraphics[width=15mm]{spin_schwarz}

\includegraphics[width=15mm]{spin_schwarz}
\includegraphics[width=15mm]{spin_schwarz}
\includegraphics[width=15mm]{spin_schwarz}
\includegraphics[width=15mm]{spin_schwarz}
\includegraphics[width=15mm]{spin_schwarz}
\includegraphics[width=15mm]{spin_schwarz}
\includegraphics[width=15mm]{spin_schwarz}
\includegraphics[width=15mm]{spin_schwarz}
\includegraphics[width=15mm]{spin_schwarz}
\includegraphics[width=15mm]{spin_schwarz}

\vspace*{4mm}

\includegraphics[width=24mm]{bond_blau}
\includegraphics[width=24mm]{bond_blau}
\includegraphics[width=24mm]{bond_blau}
\includegraphics[width=24mm]{bond_blau}
\includegraphics[width=24mm]{bond_blau}
\includegraphics[width=24mm]{bond_blau}
\includegraphics[width=24mm]{bond_blau}

\includegraphics[width=24mm]{bond_blau}
\includegraphics[width=24mm]{bond_blau}
\includegraphics[width=24mm]{bond_blau}
\includegraphics[width=24mm]{bond_blau}
\includegraphics[width=24mm]{bond_blau}
\includegraphics[width=24mm]{bond_blau}
\includegraphics[width=24mm]{bond_blau}

\includegraphics[width=24mm]{bond_blau}
\includegraphics[width=24mm]{bond_blau}
\includegraphics[width=24mm]{bond_blau}
\includegraphics[width=24mm]{bond_blau}
\includegraphics[width=24mm]{bond_blau}
\includegraphics[width=24mm]{bond_blau}
\includegraphics[width=24mm]{bond_blau}

\includegraphics[width=24mm]{bond_blau}
\includegraphics[width=24mm]{bond_blau}
\includegraphics[width=24mm]{bond_blau}
\includegraphics[width=24mm]{bond_blau}
\includegraphics[width=24mm]{bond_blau}
\includegraphics[width=24mm]{bond_blau}
\includegraphics[width=24mm]{bond_blau}

\includegraphics[width=24mm]{bond_blau}
\includegraphics[width=24mm]{bond_blau}
\includegraphics[width=24mm]{bond_blau}
\includegraphics[width=24mm]{bond_blau}
\includegraphics[width=24mm]{bond_blau}
\includegraphics[width=24mm]{bond_blau}
\includegraphics[width=24mm]{bond_blau}

\includegraphics[width=24mm]{bond_blau}
\includegraphics[width=24mm]{bond_blau}
\includegraphics[width=24mm]{bond_blau}
\includegraphics[width=24mm]{bond_blau}
\includegraphics[width=24mm]{bond_blau}
\includegraphics[width=24mm]{bond_blau}
\includegraphics[width=24mm]{bond_blau}

\includegraphics[width=24mm]{bond_rot}
\includegraphics[width=24mm]{bond_rot}
\includegraphics[width=24mm]{bond_rot}
\includegraphics[width=24mm]{bond_rot}
\includegraphics[width=24mm]{bond_rot}
\includegraphics[width=24mm]{bond_rot}
\includegraphics[width=24mm]{bond_rot}

\includegraphics[width=24mm]{bond_rot}
\includegraphics[width=24mm]{bond_rot}
\includegraphics[width=24mm]{bond_rot}
\includegraphics[width=24mm]{bond_rot}
\includegraphics[width=24mm]{bond_rot}
\includegraphics[width=24mm]{bond_rot}
\includegraphics[width=24mm]{bond_rot}

\includegraphics[width=24mm]{bond_rot}
\includegraphics[width=24mm]{bond_rot}
\includegraphics[width=24mm]{bond_rot}
\includegraphics[width=24mm]{bond_rot}
\includegraphics[width=24mm]{bond_rot}
\includegraphics[width=24mm]{bond_rot}
\includegraphics[width=24mm]{bond_rot}

\newpage

\includegraphics[width=18.9cm]{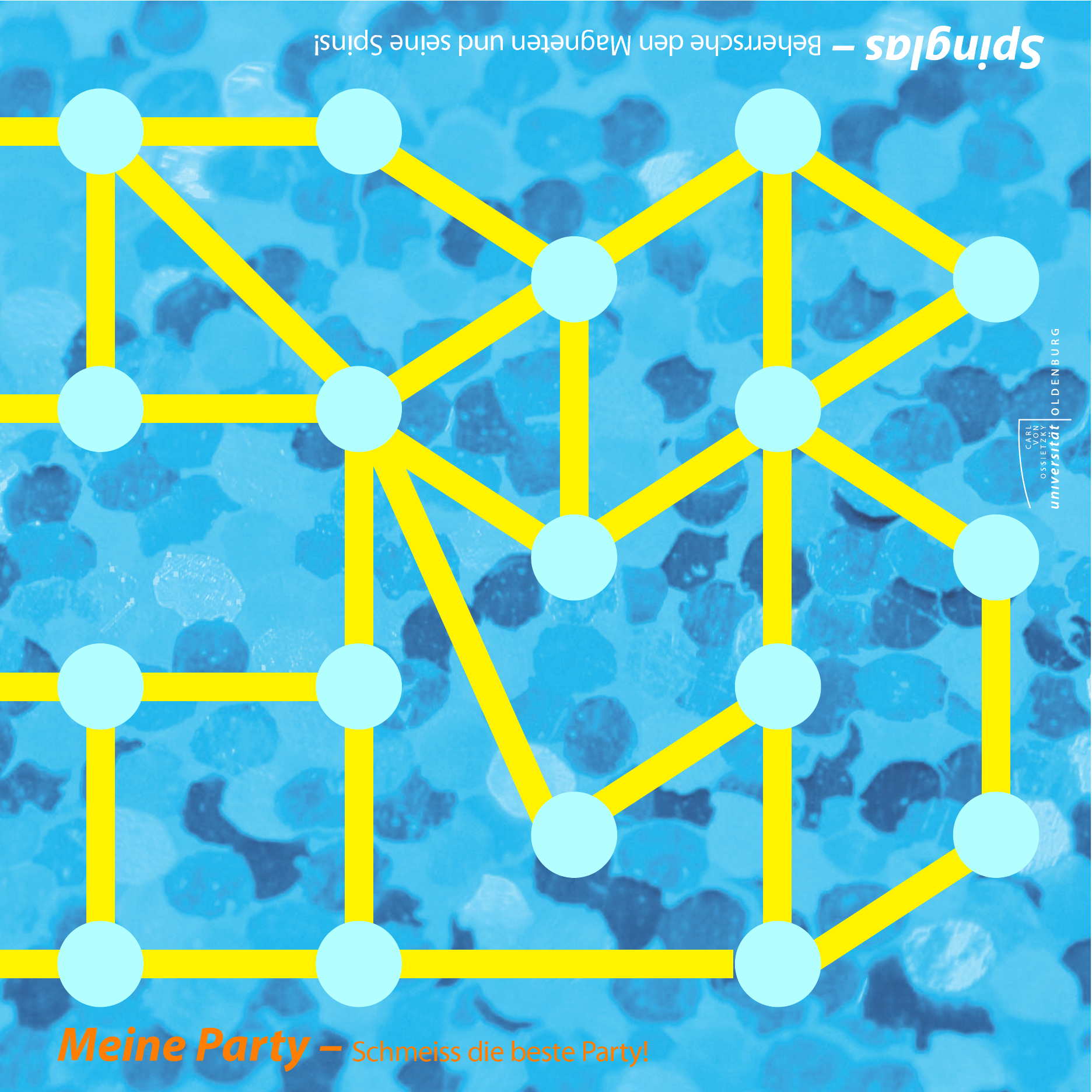}

\newpage

(print this page twice)\\

\begin{minipage}[t]{5cm}
\vspace*{30mm}

\includegraphics[width=\textwidth]{karte_spezial_final}
\end{minipage}
\begin{minipage}[t]{5cm}
\vspace*{30mm}

\includegraphics[width=\textwidth]{karte_master_final}
\end{minipage}
\begin{minipage}[t]{5cm}
\vspace*{30mm}

\includegraphics[width=\textwidth]{karte_kontakt_final}
\end{minipage}

\vspace*{-20mm}

\begin{minipage}[t]{5cm}
\vspace*{30mm}

\includegraphics[width=\textwidth]{karte_entferne_final}
\end{minipage}
\begin{minipage}[t]{5cm}
\vspace*{30mm}

\includegraphics[width=\textwidth]{karte_doppel_final}
\end{minipage}
\begin{minipage}[t]{5cm}
\vspace*{30mm}

\includegraphics[width=\textwidth]{karte_feld_final}
\end{minipage}

\end{document}